# Future prospect of anisotropic 2D tin sulfide (SnS) for emerging electronic and quantum device applications

Abdus Salam Sarkar

Department of Physics, Stevens Institute of Technology, Hoboken, NJ, 07030, USA

Center for Quantum Science and Engineering, Stevens Institute of Technology, Hoboken, NJ, 07030, USA

Email: asarkar5@stevens.edu; salam.phys@gmail.com

## Abstract

The family of anisotropic two-dimensional (2D) emerging materials is rapidly evolving due to their low crystal symmetry and in-plane structural anisotropy. Among these, 2D tin sulfide (SnS) has gained significant attention because of its distinctive crystalline symmetry and the resulting extraordinary anisotropic physical properties. This perspective explores recent developments in anisotropic 2D SnS. In particular, it highlights advances in isolating high-quality SnS monolayers (1L-SnS) and in applying advanced techniques for anisotropic characterization. The discussion continues with an overview of the anisotropic optical and electronic properties of SnS, followed by recent progress in emerging electronic device applications, including energy conversion & storage, neuromorphic (synaptic) systems, spintronics and quantum technologies. In addition to presenting significant research findings on SnS, this perspective outlines current limitations and discusses emerging opportunities and future prospects for its application in quantum devices.

## 1. Introduction

Anisotropy in materials introduces a rich and versatile dimension to advancements in materials science, particularly in emerging semiconductor device technologies (Li et al., 2019b; Xia et al., 2019; Sarkar and Stratakis, 2020; Y.-R. Chang et al., 2023; Zhitao Lin et al., 2025). Among these, two-dimensional (2D) anisotropic materials have gained prominence in quantum materials research due to their extreme quantum confinement at the atomic scale (Dwyer et al., 2019; Sarkar et al., 2021). The inherent anisotropy of certain 2D materials gives rise to extraordinary quantum phenomena, unlocking new possibilities for advanced applications. Recent developments in metal monochalcogenides (Sarkar and Stratakis, 2020; Sarkar et al., 2020, 2023), chemical formula (MX, M=metal, X=Chalcogens) have positioned them at the forefront in anisotropic materials research. In particular, 2D tin sulfide (SnS) has attracted significant attention due to its unique direction-dependent properties. This material has been studied from various perspectives, including its electronic, optical, and excitonic behaviors, which are crucial for next-generation energy, optoelectronic, spintronics and quantum devices.

In 2015, the first isolation of few-layer SnS revealed its exotic optical and electronic properties (Brent et al., 2015). The SnS exhibits either a direct or indirect band gap and its polarization dependent anisotropic optical and electronic behaviors have been actively explored (Hanakata et al., 2016; Tian et al., 2017; Chen et al., 2018; Bao et al., 2019; Li et al., 2019a; Gomes and Carvalho, 2020; Higashitarumizu et al., 2020; Kwon et al., 2020; Gao et al., 2022; Zi et al., 2022; Y.-R. Chang et al., 2023). The anisotropic nature of electronic transport along the armchair (AC) and zigzag (ZZ) directions gives rise to distinct features. Direction dependent nonlinear optical and valley-related properties, including multiferroicity, have also been investigated (Hanakata et al., 2016; Rodin et al., 2016; Wang and Qian, 2017; Chen et al., 2018; Maragkakis et al., 2022; Verma et al., 2023; Sutter et al., 2024). Monolayer SnS adopts an orthorhombic crystal structure with a *Pnma* space group (**Figure 1a-c**). The structure exhibits low crystal symmetry ($C_{2v}$) and broken inversion symmetry. These unique characteristics enable the material to support a range of novel order parameters, including optical nonlinearity, spin-orbit coupling, ferroelectricity, polarization, and synaptic properties.

A variety of exotic phenomena have been predicted or demonstrated in monolayer SnS, such as the bulk photovoltaic effect (Y.-R. Chang et al., 2023; N. Kumar et al., 2025), synaptic behavior for neuromorphic computing (Kwon et al., 2020; Khot et al., 2024), second- and third-order optical nonlinearity (Maragkakis et al., 2022; G. M. Maragkakis et al., 2024), spin-valley polarization (Hanakata et al., 2016; Rodin et al., 2016; Chen et al., 2018; Lin et al., 2018; A. K. Tołłoczko et al., 2025), ferroelectricity (Bao et al., 2019; Higashitarumizu et al., 2020; Kwon et al., 2020; Y.-R. Chang et al., 2023), giant piezoelectricity (Khan et al., 2020; Cao et al., 2021; Yoo et al., 2023), multiferroicity (Wang and Qian, 2017), chiral phonons (Skelton et al., 2017), thermoelectric effects (Asfan diyar et al., 2024; Feng-ning Xue et al., 2025), spintronics (H. Yang et al., 2024), superconductivity (Jianyong Chen et al., 2025), and quantum confinement effects (Dwyer et al., 2019). Additionally, the strong second- and third- harmonic generation observed in mono- and few- layer SnS holds great promise for nonlinear optoelectronics and the generation of entangled photon pairs (EPPs) via spontaneous parametric down-conversion (SPDC) for quantum technological applications (A. S. Sarkar, 2025).

This perspective covers the isolation of anisotropic 2D SnS monolayers, their physical properties, advanced characterization techniques, and emerging electronic device applications, including those in quantum technology. More specifically, the isolation of single-layer SnS using conventional methods is explored, along with microscopy and spectroscopic characterizations.

Furthermore, applications in energy, optoelectronic, neuromorphic computing, spintronic, and quantum devices are discussed. The perspective also highlights current limitations and explores emerging opportunities for future device applications.

## 2. Synthesis characterizations and properties

Anisotropy refers to the property of a material exhibiting different physical characteristics depending on the direction of measurement. In particular, structural anisotropy at the crystal level plays a pivotal role in this behavior. The concept is closely related to asymmetry in materials or crystallographic structures, and anisotropy typically arises from differences along specific crystallographic axes. In this context, it discusses the polarization mechanisms in the anisotropic material SnS. For instance, the polarization-dependent optical properties arising from its intrinsic anisotropy. The polarization-dependent optical properties originate from their fundamental physical nature (Sharma et al., 2018). These can generally be classified into three categories: (1) anisotropy due to structural asymmetry, (2) the presence of inequivalent electronic valleys, and (3) interactions among quasiparticles, such as excitons and polarized plasmons or photons.

To understand and describe polarization mechanisms in anisotropic materials, the semiclassical Drude model is employed. In a simplified model for 2D layered materials, polarization contribution along one direction is often considered homogeneous, while the polarization responses in the other two directions can be characterized using the relative permittivity ($\epsilon_{jj}$) and optical conductivity ($\sigma_{jj}$), as described by

$$\epsilon_{jj} = \epsilon_r + \frac{i\sigma_{jj}}{\epsilon_0 \omega t} \quad (1)$$

$$\sigma_{jj} = \begin{bmatrix} \sigma_{xx} \; \sigma_{xy} \\ \sigma_{yx} \; \sigma_{yy} \end{bmatrix} \quad (2)$$

j=$x$, $y$ (considering $x$-$y$ plane of the material). Here, $\epsilon_0$ and $\epsilon_r$ represent the vacuum permittivity and the relative permittivity, respectively. $\omega$ and t denote the optical frequency and the thickness of the layered material. The polarization-dependent optical and electronic properties of anisotropic layered materials are governed by these parameters. In anisotropic 2D SnS, the atomic arrangement within the crystal leads to highly direction-dependent (anisotropic) carrier effective masses in real space (**Figure 1a-c**). Specifically, anisotropic 2D materials with linear electronic dispersion exhibit an optical conductivity tensor where $\sigma_{xx} \neq \sigma_{yy}$ and $\sigma_{xy}=\sigma_{yx}=0$, while isotropic materials follow $\sigma_{xx}=\sigma_{yy}$ and $\sigma_{xy}=\sigma_{yx}=0$.

Anisotropy in the physical properties of SnS has been extensively investigated through both theoretical and experimental approaches. For instance, anisotropic absorption and photoluminescence, including nonlinear optical (NLO) properties of SnS, have been explored (Batool et al., 2022). These studies reveal that the optical properties of SnS are directionally dependent. Notably, second- and third-harmonic generation (SHG and THG) along the AC and ZZ directions exhibit significant differences. The average THG anisotropy in SnS is estimated to be 0.75 ± 0.22 (G. M. Maragkakis et al., 2024). In contrast, the anisotropic SHG response of a SnS layer has also been reported. The anisotropic nature of these NLO properties originates from the inherent crystal symmetry of SnS.

In addition to optical anisotropy, anisotropy in the electronic properties of SnS plays a critical role in device applications. Specifically, the anisotropic ratio between the AC and ZZ directions presents opportunities for novel functionalities. The effective mass ($m$) strongly influences carrier mobility (μ), as $\mu \propto 1/m$. Importantly, the effective mass along the armchair direction ($m_{AC}$) differs from that along the zigzag direction ($m_{ZZ}$), which contributes significantly to the electronic anisotropy of SnS. For example, the experimentally observed anisotropy ratio of SnS is approximately 1.7, closely matching the theoretically predicted values (Vidal et al., 2012; Guo et al., 2015). Moreover, the theoretical aspects of anisotropy have also been investigated. For instance, Sandonas et al. (Medrano Sandonas et al., 2016) theoretically estimated the anisotropic thermoelectric response in SnS. A significantly high figure of merit (ZT) was obtained at room temperature, reaching 1.6 and 0.95 along the AC and ZZ directions, respectively. These predicted ZT values are in close agreement with the experimental ZT value of approximately 0.7 (Yixuan Hu et al., 2024).

The synthesis of electronic grade materials is a critical first step toward unlocking the rich physics and exceptional properties required for high performance device applications. Since the first isolation of SnS via liquid phase exfoliation (LPE) (Brent et al., 2015; Sarkar et al., 2020, 2023; Sarkar and Stratakis, 2021), various synthesis methodologies have been explored to produce few- and mono- layer SnS (FL/1L-SnS) structures. Conventional LPE methods have been effective in isolating FL-SnS sheets, while non-conventional approaches such as thermally assisted LPE (T-LPE) have been developed to overcome the strong interlayer binding energy, (Sarkar et al., 2023) enabling the 1L-SnS (**Figure 1e**). However, the lateral dimensions of these monolayers are typically limited to the micrometer scale due to the buckled crystal structure of SnS and the non-uniform shear forces inherent in the exfoliation process. In contrast, vapor-phase techniques such as chemical vapor deposition (CVD) have demonstrated significant potential in producing ultrathin SnS layers with improved crystallinity and uniformity. Monolayer and few-layer SnS have been successfully synthesized using these method (**Figure 1d**) (Kazuki Koyama et al., 2025), although challenges remain in achieving consistent layer thickness and large area coverage. The anisotropic crystal structure of SnS, coupled with the differing electron affinities of its constituent atoms, further complicates the growth of high-quality monolayers, often resulting in the presence of crystal defects and residual impurities.

Among all techniques, mechanical exfoliation (ME) of bulk crystals continues to produce the highest purity, electronic grade monolayers. Scotch tape is primarily used to break the weak van der Waals bonds between layers for mechanical exfoliation. However, its low exfoliation yield and poor reproducibility render it a non-conventional method. In the case of SnS, the stronger interlayer binding energy compared to conventional van der Waals materials such as TMDs makes mechanical exfoliation particularly challenging. Recent advances in metal assisted mechanical exfoliation (M-ME) have shown notable improvements in yield and layer uniformity. Notably, Rao co-workers demonstrated a M-ME approach that successfully produced monolayer SnS with high structural integrity (Nicolas Gauriot et al., 2025). Other emerging methods such as chemical synthesis (Li et al., 2019), liquid metal (Krishnamurthi et al., 2020), Plasma-induced (Kim et al., 2018), and laser-induced synthesis (A. V. Averchenko et al., 2024) are also being actively explored to produce high-quality SnS thin layers. These approaches continue to evolve, aiming to optimize both scalability and material performance for integration into next-generation quantum and optoelectronic devices.

The synthesized anisotropic SnS structures have been critically investigated using various advanced characterization techniques. Initial exfoliation efforts were primarily limited to obtaining few-layer SnS. However, M-ME successfully overcame the strong interlayer forces, enabling isolation down to the monolayer level (**Figure 1g**). CVD technique by contrast (Sutter et al., 2020; Zhang et al., 2023; Kazuki Koyama et al., 2025), have generally been limited to producing multilayer films (**Figure 1h**), as the inherent anisotropy in atomic arrangement restricts the nucleation of tin and sulfur atoms along the AC/ZZ directions. Zo et al. (Kazuki Koyama et al., 2025) addressed this challenge by varying the sulfur vapor concentration relative to tin, using high purity elemental precursors. As a result, they achieved the growth of monolayer SnS crystals (**Figure 1i**) with lateral dimensions reaching several tens of micrometers. This large scale monolayer growth was enabled by carefully controlling the sublimation rate of bulk SnS crystals.

The wavy nature of the SnS crystal structure has been revealed using scanning tunneling microscopy (STM). Cross-sectional and top-view imaging distinctly highlight the anisotropic structure of monolayer SnS. Notably, the thermal expansion coefficients of SnS are significantly higher than those of conventional TMDs, which facilitates monolayer isolation via T-LPE. The resulting monolayers (**Figure 1j**) are highly crystalline and exhibit an orthorhombic crystal structure (**Figure 1k**) (Sarkar et al., 2020). The intrinsic anisotropy along the AC/ZZ directions is further confirmed through AFM images. In addition, a variety of optics-based characterization methods have been employed to investigate the anisotropy in SnS. For example, polarized-absorption spectroscopy (Nicolas Gauriot et al., 2025), Raman spectroscopy (Tian et al., 2017), second harmonic generation (SHG) spectroscopy (Y.-R. Chang et al., 2023), and polarization-resolved optical microscopy (Zhu et al., 2021; Maragkakis et al., 2022) have been utilized to explore the unique anisotropic optoelectronic properties of SnS.

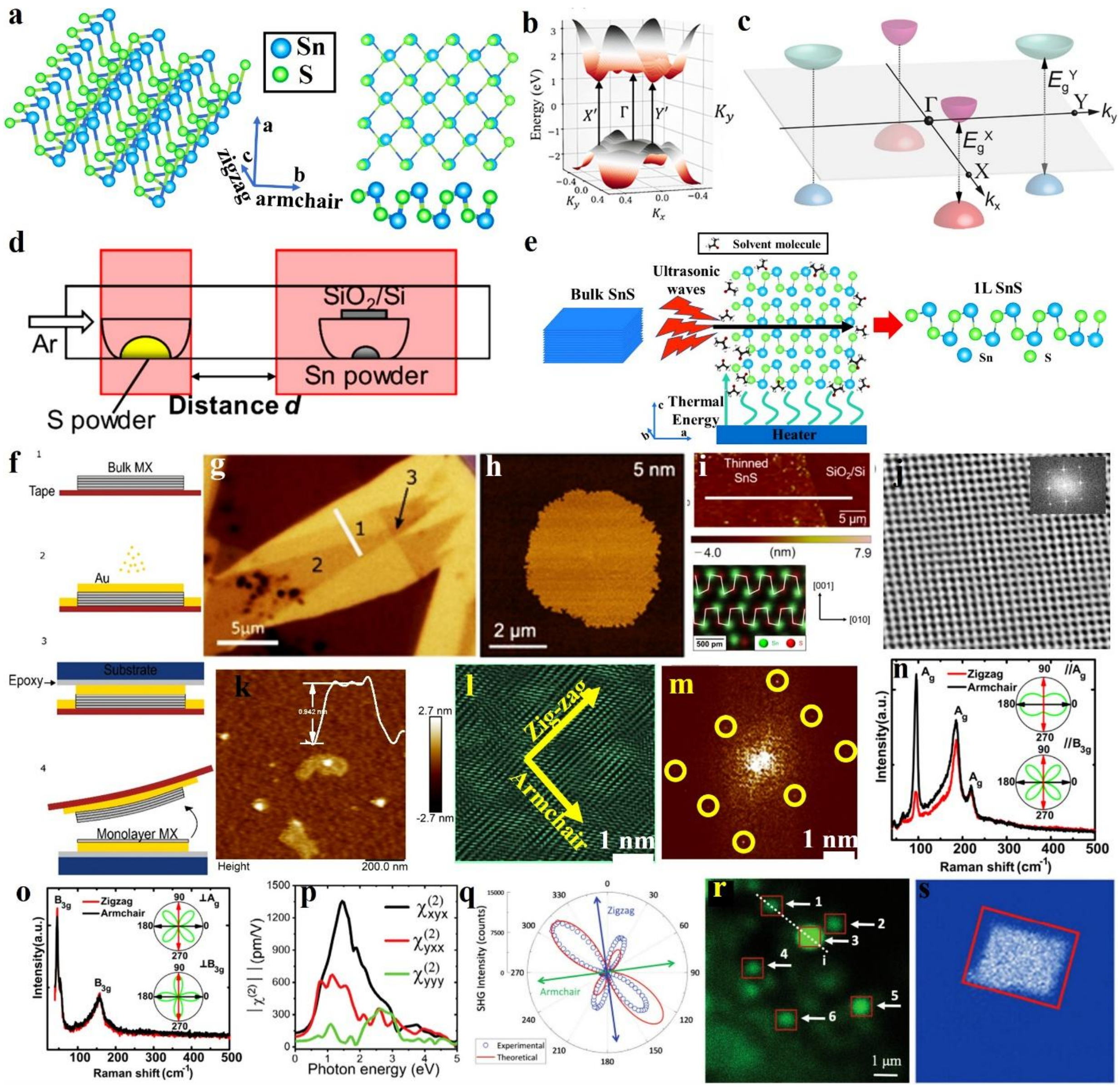

**Figure 1. Anisotropic Crystal Structure, Isolation, and Characterization of SnS** **(a)** Atomic structure of anisotropic two-dimensional SnS. *Left:* Side view of the anisotropic structure. *Right:* Top view of the SnS crystal. *Bottom right:* Monolayer of SnS. ***b*** and ***c*** is the armchair and zigzag directions of the anisotropic SnS crystal. **(b)** Electronic energy spectrum of monolayer SnS. X′ and Y′ represent the two distinct valleys. Adopted with permission (Nguyen Thanh Tien et al., 2025). Copyright 2024, American Physical Society. **(c)** Schematic illustration of the conduction and valence band edges of SnS. The band gap, $E_g$, varies with the number of layers. $E_g^x$ and $E_g^y$ represent the band gaps of the non-degenerate valleys along the armchair (x/b) and zigzag (y/c) directions, respectively. Reproduced with permission. Copyright 2024, the Royal Society of Chemistry (E. Sutter et al., 2024). **(d)** Schematic illustration of the chemical vapor deposition (CVD) system used for the growth of SnS crystals. Adapted from (Kazuki Koyama et al., 2025), Copyright 2025, American Chemical Society, under the CC BY-NC-ND 4.0 license. **(e)** Schematic representation of the thermally assisted liquid-phase exfoliation (T-LPE) process for monolayer SnS. Reproduced (Sarkar et al., 2023). Copyright 2023, Wiley-VCH under the Creative Commons CC BY license. **(f)** Schematic illustration of metal-assisted mechanical exfoliation (M-ME) used for SnS crystal preparation. **(g)** Optical micrograph of a monolayer and few-layer SnS sample obtained via M-ME. Reproduced from (Nicolas Gauriot et al., 2025). Copyright 2025, Institute of Physics under Creative Commons Attribution (CC BY). (**h)** Atomic force microscopy (AFM) image of a PVD-grown SnS flake. Reproduced with permission (Sutter et al., 2020). Copyright, 2020 American Chemical Society. **(i)** Atomic force microscopy image with a corresponding height profile along the white line of a CVD grown SnS flake, alongside a cross-sectional scanning tunneling microscope (STM) image of SnS. The positions of Sn and S atoms are identified through EDS mapping of the SnS crystal. Adapted from (Kazuki Koyama et al., 2025), Copyright 2025, American Chemical Society, under the CC BY-NC-ND 4.0 license. **(j)** Fast Fourier transform (FFT) filtered atomic-resolution image of a few-layer SnS flake obtained via liquid-phase exfoliation (LPE). The inset shows the FFT pattern of a selected region of the SnS flake. Adapted from Sarkar et al. (Sarkar et al., 2020). Copyright 2020 Springer Nature, under the Creative Commons Attribution 4.0 International License. **(k)** AFM image of a SnS monolayer obtained via T-LPE with 20 hours of ultrasonication in acetone. The inset shows the corresponding AFM height profile of the monolayer flake. **(l)** Lattice-resolution

image of the SnS flake, revealing its anisotropic crystal structure. The orthorhombic lattice with distinct armchair and zigzag directions is clearly identified. **(m)** FFT pattern of the lattice-resolution image of the SnS monolayer flake. Reproduced (Sarkar et al., 2023). Copyright 2023, Wiley-VCH under the Creative Commons CC BY license. (**n)** Parallel polarization configuration between the incident and Raman scattered light. **(o)** Perpendicular polarization configuration. The inset shows a polar plot of the Raman signal intensity as a function of the excitation laser polarization and the anisotropic (armchair) direction. Reproduced with permission (Tian et al., 2017). Copyright, 2017 American Chemical Society. **(p)** Calculated second-harmonic generation (SHG) susceptibilities along the bulk armchair direction of SnS. The spectra show $\chi$ as a function of the incident laser energy. Reproduced with permission (R. Moqbel et al., 2024). Copyright, 2024 Wiley-VCH. **(q)** Polar plot of second-harmonic generation (SHG) measured from few-layer SnS. The red solid line represents the theoretical SHG calculation. Reproduced from Wiley-VCH. Copyright CC BY-NC 4.0 (Y.-R. Chang et al., 2023). **(r)** Microscopic second-harmonic generation (SHG) measurements of SnS flakes. The SHG intensity is shown as a function of the linearly polarized excitation angle. Reproduced with permission (Maragkakis et al., 2022). Copyright, 2022, Wiley-VCH. and **(s)** SHG intensity mapping image of a monolayer SnS. Reproduced from Wiley-VCH (H. Yang et al., 2024). Copyright 2020, Creative Commons CC-BY-NC-ND license.

Anisotropic absorption spectra of SnS were acquired under illumination with linearly polarized light along the *x*- and *y*-axes. The spectra reveal distinct absorption edges corresponding to the material's anisotropic band structure, with bandgap energies determined to be 1.173 eV and 1.147 eV, respectively (Chen et al., 2018). Another key technique used to probe anisotropy in SnS is polarized Raman spectroscopy (PRS) (Cong et al., 2020; Gong et al., 2022; Zhao et al., 2024). In this method, the Raman signal intensity ($I$) is given by the expression $I \propto |e_i R e_s^T|^2$, where $e_i$ and $e_s$ are the polarization vectors of the incident and scattered light, respectively. As the SnS crystal adopts an orthorhombic structure with $C_{2v}$ point group symmetry. This symmetry gives rise to 24 phonon modes at the Brillouin zone center (Γ point), which are classified as

$$\Gamma = 4A_g + 2B_{1g} + 4B_{2g} + 2B_{3g} + 2A_u + 4B_{1u} + 2B_{2u} + 4B_{3u} \quad (3)$$

where the $A_g$, $B_{1g}$, $B_{2g}$, and $B_{3g}$ are the optically active Raman modes. Among all Raman-active modes in SnS, the $A_g$ and $B_{3g}$ modes are optically active and prominently observed in polarized Raman measurements. SnS belongs to the orthorhombic crystal system, within this symmetry, the optically active Raman tensors for the $A_g$ and $B_{3g}$ modes are well-defined and govern the PRS scattering. The intensity of these Raman modes under various polarization configurations provides insight into the crystal orientation and anisotropy of the sample. In particular, the angular dependence of the $A_{3g}$ and $B_{3g}$ mode intensities can be used to distinguish between the AC/ZZ crystallographic directions of the SnS flake.

SHG is a nonlinear optical process that occurs in a non-centrosymmetric crystal. It arises in materials where the electric polarization (P) responds nonlinearly to an applied electric field (E) and described by

$$P(t) = \epsilon_0 \chi^{(1)} E(t) + \epsilon_0 \chi^{(2)} E^2(t) + \cdots \ldots \quad (4)$$

where, $\epsilon_0$ is free space permittivity, $\chi^{(1)}$ and $\chi^{(2)}$is the linear (first order), and nonlinear (*2nd* order) susceptibility of the material, respectively. To investigate the anisotropic nature of SnS structure particularly along the AC/ZZ directions P-SHG measurements are employed. These measurements provide insight into the crystallographic orientation and symmetry-dependent nonlinear optical response of SnS.

SHG spectroscopy of SnS layers deposited on highly oriented pyrolytic graphite (HOPG) has been investigated to explore their anisotropic optical properties (R. Moqbel et al., 2024). Due to the inherently anisotropic crystal structure of SnS, SHG exhibits a strong angular dependence when excited with linearly polarized light. Chang et al. (Y.-R. Chang et al., 2023) demonstrated polarization resolved SHG (P-SHG) (**Figure 1q**), where both the incident and detected SHG signals are linearly polarized. The resulting SHG polar plots from FL-SnS showed excellent agreement with theoretical predictions, confirming the distinct optical response along the AC/ZZ crystallographic directions. Beyond conventional SHG spectroscopy, Maragakis et al. (Maragkakis et al., 2022) performed P-SHG microscopy on randomly oriented SnS flakes dispersed on a transparent substrate. **Figure 1r** presents representative P-SHG images of multiple SnS flakes within the same field of view. Each flake is labeled numerically to examine the polarization dependent anisotropic response and to identify the AC/ZZ orientations. In this study, P-SHG images were recorded as a function of the linear polarization angle ($\phi$) of the excitation laser. Regions of interest (ROIs) within individual SnS flakes reflect the angular dependence of the $\chi^{(2)}$ tensor with respect to the crystallographic axes. Consequently, the polar plots of P-SHG intensity are directly correlated with the flake's crystal orientation, allowing for precise identification of the AC/ZZ directions. Furthermore, SHG measurements from large-area monolayer SnS samples have also been employed to determine crystal orientation, as shown in **Figure 1s.** These findings reinforce the effectiveness of P-SHG as a non-invasive optical tool to probe anisotropic nonlinear responses and to resolve in-plane crystallographic axes in 2D materials.

## 3. Emerging device applications

The recent development of ultrathin SnS layers has generated significant interest due to their rich physical properties and potential for various emerging applications. In particular, the structural anisotropy arising from the puckered structure with low crystal symmetry of SnS makes it highly promising for advanced electronic and optical technologies, including next-generation quantum applications.

### 3.1 **Electronic** Device Applications

The piezoelectric and ferroelectric properties arising from broken crystal symmetry in anisotropic 2D SnS have positioned it as a promising material for energy harvesting and computing applications (Cao et al., 2021; Yoo et al., 2023; Y.-R. Chang et al., 2023; C. Chen et al., 2025). In particular, the bulk photovoltaic effect (BPVE) in ferroelectric materials is gaining significant attention. This effect allows for spontaneous polarization reversal in non-centrosymmetric crystals, a phenomenon known as domain wall photovoltaic effect (DW-PVE). Its potential lies in enabling the Shockley-Queisser limit to be surpassed in single-junction photovoltaics. For example, Chang et al. (Y.-R. Chang et al., 2023) have directly observed shift current and ferroelectric domains in SnS PV. The photocurrent line profiles along the AC direction revealed distinct photocurrent peaks (**Figure 2a, b**). These current peaks were corroborated by photocurrent mapping, which is dependent on the ferroelectric domain structure. The ferroelectric domains in SnS enhance the pathways for photodetection and energy conversion, offering superior performance compared to other 2D materials (**Figure 2c**).

Notably, SnS exhibits layer number dependent ferroelectric responses, primarily influenced by crystal layer stacking (**Figure 2e**). Similarly, SHG intensity and the on/off current ratio are also dependent on the number of layers. Nonetheless, the polarization switching in SnS occurs through a structural transition from a puckered geometry to a cubic NaCl like phase. Remarkably, SnS demonstrates a strong anisotropic BPVE effect. **Figure 2f** shows the anisotropic DW-PVE behavior of an SnS device along its anisotropy directions (R. Nanae et al., 2024). Photocurrent measurements were conducted using 488 nm polarized light, with the response originating from the device's Schottky junction. At zero bias voltage, the I-V characteristics are nearly identical in both directions, while the dark current remains negligible. This anisotropic response is attributed to the influence of 90° ferroelectric domains and/or the DW-PVE, as illustrated in the inset of **Figure 2f**.

In addition to the BPVE observed in single ferroelectric domains of anisotropic SnS, other energy conversion applications have become major areas of focus. These applications are evolving rapidly due to their potential in emerging technologies. Anisotropic 2D materials have recently garnered significant attention owing to their rich physical properties, particularly their directional dependent transport behavior. For instance, their anisotropic optical responses, electrical conductivity, and high physical anisotropy ratios make them stand out compared to conventional 2D materials. The electrocatalytic production of formate from $CO_2$ offers a promising pathway for sustainable energy generation, making it a key technology for future energy systems. Various nanomaterials have been investigated to uncover the fundamental mechanisms and explore practical applications (Y. Liu and C. Huang, 2025). Notably, Zou et al. (Zou et al., 2021) studied anisotropic 2D SnS and demonstrated the dramatic effect of electrolyte alkalinity in expanding the potential window for $CO_2$ electroreduction. The SnS nanosheets achieved a remarkably high Faradaic efficiency of 88% for formate production at a current density of 120 mA/cm$^2$. This outstanding performance is attributed to the interplay of three competing reactions in anisotropic 2D SnS: the evolution of formate, $H_2$, and CO (**Figure 2g**). Moreover, the phonon transport plays a pivotal role in determining the efficiency of thermal energy conversion, making the study of thermoelectric properties essential for device applications. A recent study (Feng-ning Xue et al., 2025) investigated the thermoelectric transport behavior of SnS, including the Seebeck coefficient

(S) and electrical conductivity (σ). The power factor (PF) is defined as PF = $S^2\sigma$, and the thermoelectric figure of merit is given by ZT (**Figures 2h and 2i**). At 300 K, the PF of SnS was estimated to be 0.11 mW/cm·$K^2$ and 0.15 mW/cm·$K^2$ for p-type and n-type doping, respectively (**Figure 2i**). The maximum ZT values in the in-plane (xy) direction were reported as 0.42 and 0.46 for p-type and n-type doping, respectively. These enhanced ZT values are largely attributed to the inherently low thermal conductivity of SnS along the out-of-plane ($z$) direction.

### 3.2 Optical Device Applications

The crystal symmetrical characteristics of the band structure play a crucial role in determining the optoelectronic behavior of SnS devices. For example, ZZ-polarized light can activate the A exciton while leaving the B exciton dark. Conversely, AC-polarized light induces the B exciton while keeping the A exciton inactive (Nguyen Thanh Tien et al., 2025). This distinct polarization-dependent excitonic behavior presents exciting opportunities for quantum switching and communication technologies. Unlike the hexagonal lattices such as TMDs that exhibit valley selective circular dichroism, the anisotropic tin sulfide revealed the extraordinary and unique linearly polarized optical selectivity (Rodin et al., 2016; Chen et al., 2018; Lin et al., 2018). In fact the anisotropic nature of the band structures make them accessible by employing the linearly polarized light. The valley selectivity via photoluminescence is

$$P_{2D} = \frac{I\sigma_- - I\sigma_+}{I\sigma_- + I\sigma_+} \quad (5)$$

where $\sigma_{+/-}$ is the PL intensities, which are left/right circularly polarized light, respectively. The direct access and the identification of the valleys presence in the layered SnS has significantly strong dichoric anisotropy of upto 600%. The strong polarization degrees of 95% has been recorded (**Figure 2j, k**).

As an emerging technology, synaptic devices are evolving rapidly (Kwon et al., 2020; Hou et al., 2021; Khot et al., 2024; Tao Zhang et al., 2024; C. Chen et al., 2025). These devices emulate the function of biological synapses (**Figure 2l, m**), with artificial electronic components designed to store and process information similarly to the human brain. This concept underpins the development of neuromorphic computing systems, which aim to achieve energy efficient and highly parallel information processing. For example, the human eye responds to optical signals by converting them into electrical impulses, which are transmitted to the brain through biological synapses for visual processing. In more detail, signal transmission across a biological synaptic cleft occurs via neurotransmitters such as acetylcholine (ACh), as illustrated in **Figure 2m**. SnS devices have demonstrated excellent figures of merit for synaptic applications. For instance, Kawon et al. (Kwon et al., 2020) introduced ferroelectric analog synaptic devices based on 2D SnS. Notably, SnS device exhibited strong in-plane ferroelectric responses and synaptic behavior at room temperature. The on/off conductance ratio between the maximum ($G_{max}$) and minimum ($G_{min}$) values, along with long term plasticity (LTP/LTD), in SnS-based synaptic devices with a metal/ferroelectric semiconductor/metal (Pt/SnS/Pt) structure is crucial for reducing power consumption. The SnS based device exhibited a high incremental voltage spike response, a $G_{max}$/ $G_{min}$ ratio of ~20.5, and linear LTP/LTD behavior (**Figure 2l**), with over 100 distinct conductance states. Interestingly, multilevel conductance states were observed in the multidomain structure of the SnS device (N. Kumar et al., 2025). A remarkably stable LTP/LTD transition with uniform conductance modulation over 10,000 incremental spikes was recorded.

In another device configuration, synaptic functionality was realized using two asymmetric metal electrodes (Khot et al., 2024)The synaptic potentiation and depression characteristics were reproduced over multiple consecutive cycles. Notably, a significant

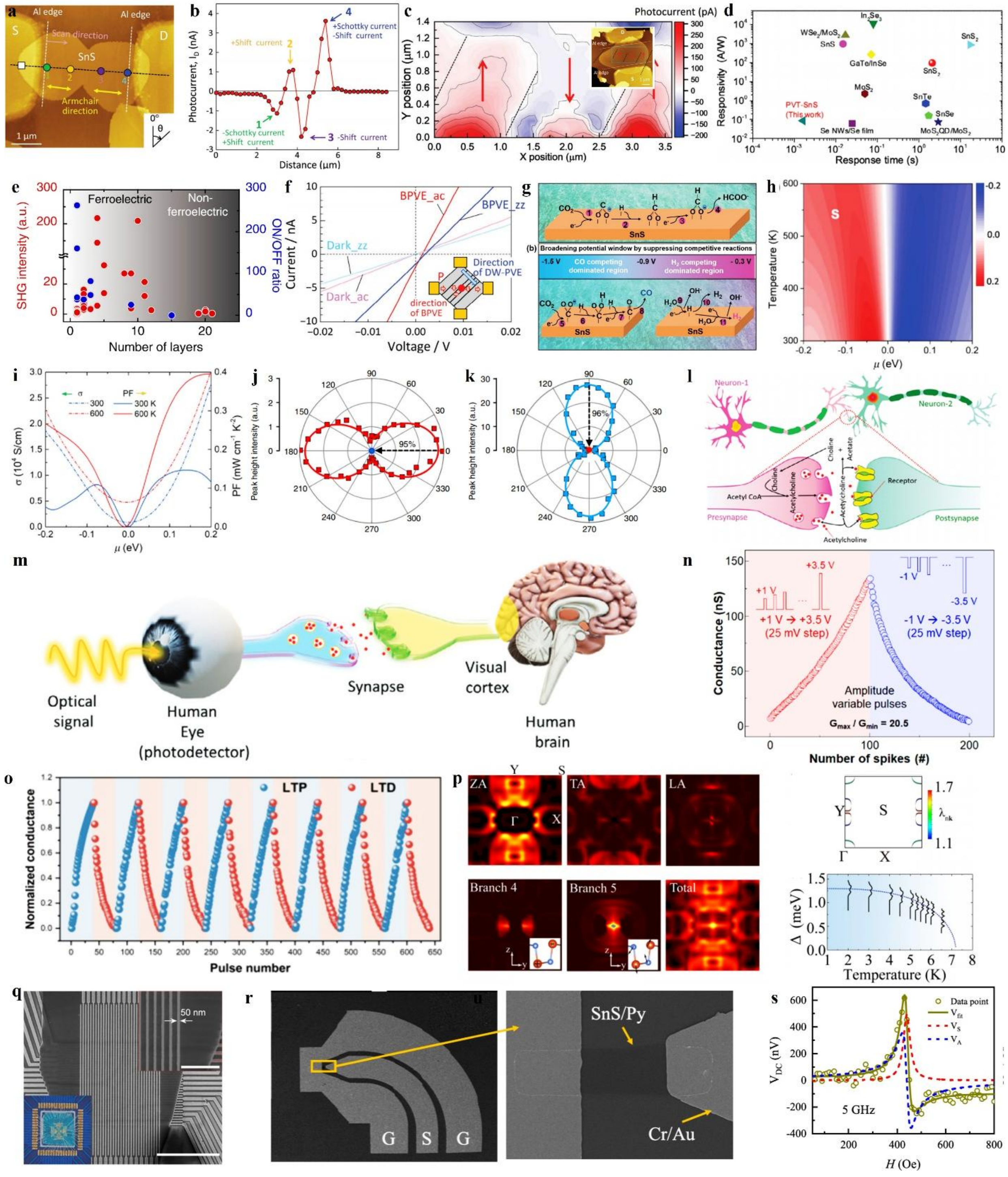

**Figure 2. Emerging applications of anisotropic 2D SnS. (a)** Atomic force microscopy (AFM) image of the fabricated SnS device. **(b)** Photocurrent characteristics of the device measured at zero drain and back gate voltage. The current was recorded along the armchair direction, indicated by the black dotted line in **a**. The positions corresponding to the photocurrent peaks are marked with matching colors in **b**, with the origin of each peak annotated. **(c)** Photocurrent mapping showing ferroelectric domain boundaries, indicated by dotted lines at regions of zero photocurrent. The domain boundaries are aligned parallel to the facet of the SnS crystal. The inset displays an AFM image of the SnS device used for the photocurrent measurements, taken in the region marked on the map. Reproduced from Wiley-VCH (Y.-R. Chang et al., 2023). Copyright CC BY-NC 4.0. **(d)** An overview of the current photodetection performance of SnS, focusing on responsivity and response time, with a comparison to other 2D materials. Adopted with permission (N. Kumar et al., 2025). Copyright 2025 Wiley-VCH. **(e)** Thickness dependence of SHG intensity and ON/OFF ratio in SnS flakes. Each data point represents an individual SnS flake. Adopted from Springer Nature (Higashitarumizu et al., 2020). Copyright 2020, Creative Commons Attribution 4.0 International License. **(f)** I–V characteristics of a SnS flake measured in anisotropic device structures along the armchair and zigzag directions. The inset shows the domain structure of the crystal. Adopted from Wiley-VCH (R. Nanae et al., 2024). Copyright, 2025, Creative Commons CC-BY-NC-ND license. **(g),** Schematic representation of the proposed mechanism for electrochemical $CO_2$ reduction. Top: Formation of formate. Bottom: $H_2$ and $CO_2$ evolution on SnS/GDL. Adopted from Wiley-VCH (Zou et al., 2021). Copyright 2021, Creative Commons CC BY license. **(h)** Mapping of the Seebeck coefficient (S) as a function of temperature (T) and chemical potential (μ). **(i)** Estimated electrical conductivity and power factor as functions of μ at 300 K and 600 K, respectively. Adopted with permission (Feng-ning Xue et al., 2025). Copyright American Physical Society 2025. **(j, k)** The polarization dependence is analyzed using PL intensity polar plots. The PL measurements were performed under polarized conditions for the valleys at 817 nm and 995 nm. The corresponding band structures are shown in ***Figure 1c***. The black dashed

arrows indicate data collected before and after a 90° rotation of the polarizer relative to the incident light. A significant reduction in PL intensities 95% at the ΓY valley and 96% at the ΓX valley was observed. Adopted from Springer Nature (Lin et al., 2018). Copyright 2018, Creative Commons Attribution 4.0 International License. **(l)** Transfer mechanism of chemical signals via the neurotransmitter (NT) acetylcholine (ACh) across biological synapses. Reproduced with permission (N. Kumar et al., 2025). Copyright, Wiley-VCH, 2025. **(m)** Object detection in biological systems by the human brain. Optical signals are received and transmitted via neurons and nerves to the brain, where they are processed for object perception. **(n)** Conductance changes in the SnS device under potentiation and depression conditions. The measurements were taken at the postsynaptic neuron using a read voltage of 0.1 V. Adopted with permission (Kwon et al., 2020). Copyright, American Chemical Society 2020. **(o)** The cyclic LTP/LTD characteristic curve of the IGZO/SnO/SnS synaptic device for neuromorphic computing. The synaptic response is triggered by 40 pulses at a frequency of 1 Hz, with a pulse width of 0.5 s. Reproduced with permission (Tao Zhang et al., 2024). Copyright, 2024 American Chemical Society. **(p)** Distribution of $\lambda_q$ for the five phonon branches under hole doping. The three active acoustic branches are ZA, TA, and LA. The inset shows the vibrational patterns of optical branches 4 and 5, where '+' and '−' indicate atomic displacements along the x-axis of the crystal. Top right: $E_g$-$k$ spectra of the electron-phonon coupling (EPC) strength $\lambda_{nk}$ at the Fermi surface under hole doping. Bottom right: Temperature dependence of the anisotropic superconducting gap in hole-doped SnS. The dashed curve serves as a visual guide. Reproduced with permission from (Jianyong Chen et al., 2025). Copyright 2025 American Physical Society. **(q)** Scanning electron microscopy (SEM) image of a fabricated high-density memristive crossbar array with a 50 nm feature size. The scale bar represents 1 μm. The inset shows an optical image of the device. Adopted with permission (Lu et al., 2021). Copyright, 2021, American Chemical Society. **(r)** SEM image of the fabricated SnS/Py spin-torque ferromagnetic resonance (STFMR) device and **(s)** ST-FMR spectra of the device measured at 5 GHz. The symmetric ($V_S$) and antisymmetric ($V_A$) components of the rectified voltage are indicated by the red and blue dashed lines, respectively. Reproduced with permission (Bangar et al., 2023). Copyright 2023, American Physical Society.

nonlinearity (NL) in potentiation and depression was observed, with values ranging between 0.3 and 0.5, respectively. On the other hand, Ye and co-workers (Tao Zhang et al., 2024) explored a reconfigurable, all-optically controlled synaptic device based on SnS and its heterostructures. In particular, the IGZO/SnO/SnS heterostructure device integrates sensing, storage, and processing functions. When integrated into an artificial neural network (ANN), the device undergoes LTP and LTD measurements over multiple cycles, which includes 40 excitatory and inhibitory conductance states (**Figure 2o**), corresponding to the synaptic weights of neural connections.

### 3.3 Quantum Device Applications

Quantum phenomena in 2D materials, magnets, topological insulator including 2D SnS, have rapidly evolved for next-generation quantum device applications. For instance, the superconductivity in 2D materials is a key aspect of advanced quantum technologies (Haoran Ji et al., 2024; Wang et al., 2024; Yichen Liu et al., 2024). In particular, the coexistence of ferroelectricity and superconductivity has emerged as a frontier in condensed matter research. Of special interest is the realization of this coexistence within a single-phase 2D material. While such phenomena have traditionally been limited to insulating systems, atomically thin layers are now being actively explored as potential platforms for the simultaneous presence of ferroelectric and superconducting properties. For example, Zhang and co-workers (Jianyong Chen et al., 2025) demonstrated the coexistence of robust ferroelectricity and superconductivity in a 2D material. This was achieved by hole doping a prototypical 2D ferroelectric insulator monolayer SnS which possesses antibonding states. Specifically, introducing approximately 0.30 holes per formula unit shifts the Fermi level into the occupied antibonding bands, effectively transforming the system into a metallic state. Remarkably, this ferroelectric metal exhibited superconductivity with a pronounced transition temperature of approximately 7 K (**Figure 2p**). The superconducting state is attributed to strong Fermi surface nesting induced by hole doping, combined with the softening of the out-of-plane acoustic phonon mode, which couples efficiently with itinerant electrons.

Recent developments in 2D SnS have enabled its integration into emerging nanoscale, high-performance electronic applications. For example, a 32 × 32 high-density memristive crossbar array with a 50 nm feature size has been successfully demonstrated (**Figure 2q**). In another promising direction, modern technology is shifting from charge-based to spin-based electronics known as spintronics which offers high speed, enhanced endurance, and reduced power consumption. These characteristics are essential for next-generation electronic and quantum technologies. Recently, Banger et al. (Bangar et al., 2023) explored the potential of SnS for spintronic applications. Specifically, they investigated spin pumping as a function of SnS thickness. Spin-torque ferromagnetic resonance (ST-FMR) measurements were performed on SnS (1.3 nm)/Py heterostructures at a radio frequency of 5 GHz (**Figures 2r and 2s**). Theoretical fitting of the data revealed the presence of an unconventional spin–orbit torque (SOT) in the SnS/Py system, which is primarily attributed to the unique crystal symmetry of SnS.

## 4. Summary & future prospect

In summary, this perspective focuses on anisotropic 2D tin sulfide. In particular, it discussed the anisotropic nature of SnS, its properties, and their fundamentals. Conventional synthesis routes along with the state-of-the-art anisotropic characterization methodologies are highlighted. Emerging electronic device applications are extensively explored, including those in energy, optoelectronics, spintronics, and next-generation quantum-based technologies.

Recent advancements have significantly improved the quality of SnS synthesized through various methods. Techniques such as M-ME, T-LPE, and CVD have made impactful contributions toward achieving ultrathin SnS layers down to the monolayer level. These thin layers exhibit remarkable anisotropic properties, opening pathways for next-generation fundamental studies and technological applications. However, the synthesis of wafer-scale, large-area, highly anisotropic monolayers remains a challenge. This may be achieved through systematic optimization of anisotropic growth along the AC and ZZ directions particularly by precisely controlling the chemical reactions involved in CVD or PVD processes.

The low crystal symmetry and intrinsic anisotropy of SnS give rise to rich electronic and optical properties, which are reflected in a wide range of emerging device applications. SnS has opened new pathways for energy conversion technologies, including the bulk photovoltaic effect, thermoelectric devices, $CO_2$ conversion, and spin-charge conversion. Moreover, SnS-based synaptic devices are playing a significant role in neuromorphic computing and artificial intelligence applications. These advancements position SnS as a promising material in the development of next-generation quantum technologies. The exceptional anisotropic fundamental properties make it highly suitable for polarized detectors, neuromorphic & synaptic devices, and spin-valley device applications.

The extraordinary properties and rich underlying physics of anisotropic SnS stem from its exceptional crystal symmetry. These unique characteristics are reflected in a range of emerging device applications, particularly in the realm of quantum technologies. For example, its pronounced nonlinear optical properties such as high SHG efficiency make it highly suitable for SPDC and the generation of EPPs. These EPPs play a pivotal role in various quantum applications, including quantum- communication, cryptography, computing, sensing, and imaging. Overall, the anisotropic nature and exceptional fundamental properties of SnS hold great promise for the development of next-generation anisotropic electronic devices, especially in the rapidly evolving field of quantum technology.

**Acknowledgements.** The author is thankful for the kind permission from the corresponding publishers or authors to reproduce the figures included in this article.

**Conflict of Interest.** The authors declare no conflict of interest

**References.**

A. K. Tołłoczko, J. Ziembicki, M. Grodzicki, J. Serafińczuk, M. Rosmus, N. Olszowska, et al. (2025). Linear Dichroism of the Optical Properties of SnS and SnSe Van der Waals Crystals. *Small* 21, 2410903. Available at: https://doi.org/10.1002/smll.202410903.

A. S. Sarkar (2025). Advancements in Entangled Photon Pairs in 2D Van der Waals Materials for On-chip Quantum Applications. *ArXiv*. Available at: https://doi.org/10.48550/arXiv.2505.09944.

A. V. Averchenko, O. A. Abbas, I. A. Salimon, E. V. Zharkova, E. D. Grayfer, S. Lipovskikh, et al. (2024). Laser-Induced Synthesis of Tin Sulfides. *Small* 20, 2401891. Available at: https://doi.org/10.1002/smll.202401891.

Asfan diyar, Wenhua Xue, Jun Mao, Kejia Liu, Qian Zhang, and Jing-Feng Li (2024). Thermoelectric Performance Enhancement in SnS Polycrystals Owing to Hole Doping Combined with Textured Microstructures. *ACS Appl. Mater. Interfaces* 16, 38073–38082. Available at: https://doi.org/10.1021/acsami.4c06851.

Bangar, H., Gupta, P., Singh, R., Muduli, P. K., Dewan, S., and Das, S. (2023). Optimization of growth of large-area SnS thin films and heterostructures for spin pumping and spin-orbit torque. *Phys. Rev. Mater.* 7. doi: 10.1103/PhysRevMaterials.7.094406

Bao, Y., Song, P., Liu, Y., Chen, Z., Zhu, M., Abdelwahab, I., et al. (2019). Gate-Tunable In-Plane Ferroelectricity in Few-Layer SnS. *Nano Lett* 19, 5109–5117. doi: 10.1021/acs.nanolett.9b01419

Batool, A., Zhu, Y., Ma, X., Saleem, M. I., and Cao, C. (2022). DFT study of the structural, electronic, and optical properties of bulk, monolayer, and bilayer Sn-monochalcogenides. *Appl. Surface Sci. Adv.* 11. doi: 10.1016/j.apsadv.2022.100275

Brent, J. R., Lewis, D. J., Lorenz, T., Lewis, E. A., Savjani, N., Haigh, S. J., et al. (2015). Tin(II) Sulfide (SnS) Nanosheets by Liquid-Phase Exfoliation of Herzenbergite: IV-VI Main Group Two-Dimensional Atomic Crystals. *J. Am. Chem. Soc.* 137. doi: 10.1021/jacs.5b08236

C. Chen, Y. Zhou, L. Tong, Y. Pang, and J. Xu (2025). Emerging 2D Ferroelectric Devices for In-Sensor and In-Memory Computing. *Adv. Mater.* 37, 2400332. Available at: https://doi.org/10.1002/adma.202400332.

Cao, V. A., Kim, M., Hu, W., Lee, S., Youn, S., Chang, J., et al. (2021). Enhanced Piezoelectric Output Performance of the SnS2/SnS Heterostructure Thin-Film Piezoelectric Nanogenerator Realized by Atomic Layer Deposition. *ACS Nano* 15, 10428–10436. doi: 10.1021/acsnano.1c02757

Chen, C., Chen, X., Shao, Y., Deng, B., Guo, Q., Ma, C., et al. (2018). Valley-Selective Linear Dichroism in Layered Tin Sulfide. *ACS Photonics* 5, 3814–3819. doi: 10.1021/acsphotonics.8b00850

Cong, X., Liu, X. L., Lin, M. L., and Tan, P. H. (2020). Application of Raman spectroscopy to probe fundamental properties of two-dimensional materials. *npj 2D Mater. Appl.* 4, 13. doi: 10.1038/s41699-020-0140-4

Dwyer, J. D., Diaz, E. J., Webber, T. E., Katzenberg, A., Modestino, M. A., and Aydil, E. S. (2019). Quantum confinement in few layer SnS nanosheets. *Nanotechnology* 30, 245705. doi: 10.1088/1361-6528/ab0e3e

E. Sutter, H. P. Komsa, and P. Sutter (2024). Valley-selective carrier transfer in SnS-based van der Waals heterostructures. *Nanoscale Horiz.* 9, 1823–1832. Available at: https://doi.org/10.1039/D4NH00231H.

Feng-ning Xue, Wei Li, Zi Li, and Yong Lu (2025). Phonon vibrational and transport properties of SnSe/SnS superlattice at finite temperatures. *Phys. Rev. B* 111, 144314. Available at: https://doi.org/10.1103/PhysRevB.111.144314.

G. M. Maragkakis, S. Psilodimitrakopoulos, L. M. A. S. S. A. L. G. K. E. S., L. Mouchliadis, A. S. Sarkar, A. Lemonis, G. Kioseoglou, et al. (2024). Anisotropic Third Harmonic Generation in 2D Tin Sulfide. *Adv. Opt. Mater.* 12, 2401321. Available at: https://doi.org/10.1002/adom.202401321.

Gao, Z. da, Jiang, Z. hui yi, Li, J. dong, Li, B. wen, Long, Y. yang, Li, X. mei, et al. (2022). Anisotropic Mechanics of 2D Materials. *Adv. Eng. Mater.* 24, 2200519. doi: 10.1002/adem.202200519

Gomes, L. C., and Carvalho, A. (2020). Electronic and optical properties of low-dimensional group-IV monochalcogenides. *J Appl Phys* 128, 121101. doi: 10.1063/5.0016003

Gong, X., Yan, T., Li, J., Liu, J., Zou, H., Zhang, B., et al. (2022). Revealing the anisotropic phonon behaviours of layered SnS by angle/temperature-dependent Raman spectroscopy. *RSC Adv.* 12, 32262–32269. doi: 10.1039/d2ra06470g

Guo, R., Wang, X., Kuang, Y., and Huang, B. (2015). First-principles study of anisotropic thermoelectric transport properties of IV-VI semiconductor compounds SnSe and SnS. *Phys. Rev. B* 92, 115202. doi: 10.1103/PhysRevB.92.115202

H. Yang, Z. Chi, G. Avedissian, E. Dolan, M. Karuppasamy, B. Martín-García, et al. (2024). Gate-Tunable Spin Hall Effect in Trilayer Graphene/Group-IV Monochalcogenide van der Waals Heterostructures. *Adv. Funct. Mater.* 34, 2404872. Available at: https://doi.org/10.1002/adfm.202404872.

Hanakata, P. Z., Carvalho, A., Campbell, D. K., and Park, H. S. (2016). Polarization and valley switching in monolayer group-IV monochalcogenides. *Phys Rev B* 94, 035304. doi: 10.1103/PhysRevB.94.035304

Haoran Ji, Yi Liu, Chengcheng Ji, and Jian Wang (2024). Two-Dimensional and Interface Superconductivity in Crystalline Systems. *Acc. Mater. Res.* 5, 1146–1157. Available at: https://doi.org/10.1021/accountsmr.4c00017.

Higashitarumizu, N., Kawamoto, H., Lee, C. J., Lin, B. H., Chu, F. H., Yonemori, I., et al. (2020). Purely in-plane ferroelectricity in monolayer SnS at room temperature. *Nat Commun* 11, 2428. doi: 10.1038/s41467-020-16291-9

Hou, Y. X., Li, Y., Zhang, Z. C., Li, J. Q., Qi, D. H., Chen, X. D., et al. (2021). Large-Scale and Flexible Optical Synapses for Neuromorphic Computing and Integrated Visible Information Sensing Memory Processing. *ACS Nano* 15, 1497–1508. doi: 10.1021/acsnano.0c08921

Jianyong Chen, Wen-Yi Tong, Ping Cui, and Zhenyu Zhang (2025). Generic approach for integrating ferroelectricity and superconductivity into a single two-dimensional monolayer. *Phys. Rev. B* 111, 094516. Available at: https://doi.org/10.1103/PhysRevB.111.094516.

Kazuki Koyama, Jun Ishihara, Nozomi Matsui, Atsuhiko Mori, Sicheng Li, Jinfeng Yang, et al. (2025). Selective Synthesis of Large-Area Monolayer Tin Sulfide from Simple Substances. *Nano. Lett.* XX, XXX–XXX. Available at: https://doi.org/10.1021/acs.nanolett.5c01639.

Khan, H., Mahmood, N., Zavabeti, A., Elbourne, A., Rahman, M. A., Zhang, B. Y., et al. (2020). Liquid metal-based synthesis of high performance monolayer SnS piezoelectric nanogenerators. *Nat. Commun.* 11, 3449. doi: 10.1038/s41467-020-17296-0

Khot, A. C., Pawar, P. S., Dongale, T. D., Nirmal, K. A., Sutar, S. S., Deepthi Jayan, K., et al. (2024). Self-assembled vapor-transport-deposited SnS nanoflake-based memory devices with synaptic learning properties. *Appl. Surface Sci.* 648, 158994. doi: 10.1016/j.apsusc.2023.158994

Kim, J. H., Yun, S. J., Lee, H. S., Zhao, J., Bouzid, H., and Lee, Y. H. (2018). Plasma-Induced Phase Transformation of SnS2 to SnS. *Sci. Rep.* 8, 10284. doi: 10.1038/s41598-018-28323-y

Krishnamurthi, V., Khan, H., Ahmed, T., Zavabeti, A., Tawfik, S. A., Jain, S. K., et al. (2020). Liquid-Metal Synthesized Ultrathin SnS Layers for High-Performance Broadband Photodetectors. *Adv. Mater.* 32, 2004247. doi: 10.1002/adma.202004247

Kwon, K. C., Kwon, K. C., Zhang, Y., Wang, L., Yu, W., Wang, X., et al. (2020). In-Plane Ferroelectric Tin Monosulfide and Its Application in a Ferroelectric Analog Synaptic Device. *ACS Nano* 14, 7628–7638. doi: 10.1021/acsnano.0c03869

Li, F., Ramin Moayed, M. M., Klein, E., Lesyuk, R., and Klinke, C. (2019a). In-Plane Anisotropic Faceting of Ultralarge and Thin Single-Crystalline Colloidal SnS Nanosheets. *Journal of Physical Chemistry Letters* 10, 993–999. doi: 10.1021/acs.jpclett.9b00251

Li, L., Han, W., Pi, L., Niu, P., Han, J., Wang, C., et al. (2019b). Emerging in-plane anisotropic two-dimensional materials. *InfoMat* 1, 54–73. doi: 10.1002/inf2.12005

Lin, S., Carvalho, A., Yan, S., Li, R., Kim, S., Rodin, A., et al. (2018). Accessing valley degree of freedom in bulk Tin(II) sulfide at room temperature. *Nat. Commun.* 9, 1455. doi: 10.1038/s41467-018-03897-3

Lu, X. F., Zhang, Y., Wang, N., Luo, S., Peng, K., Wang, L., et al. (2021). Exploring Low Power and Ultrafast Memristor on p-Type van der Waals SnS. *Nano Lett.* 21, 8800–8807. doi: 10.1021/acs.nanolett.1c03169

Maragkakis, G. M., Psilodimitrakopoulos, S., Mouchliadis, L., Sarkar, A. S., Lemonis, A., Kioseoglou, G., et al. (2022). Nonlinear Optical Imaging of In-Plane Anisotropy in Two-Dimensional SnS. *Adv. Opt. Mater.* 10, 2102776. Available at: https://doi.org/10.1002/adom.202102776.

Medrano Sandonas, L., Teich, D., Gutierrez, R., Lorenz, T., Pecchia, A., Seifert, G., et al. (2016). Anisotropic Thermoelectric Response in Two-Dimensional Puckered Structures. *J. Phys. Chem. C* 120, 18841–18849. doi: 10.1021/acs.jpcc.6b04969

N. Kumar, M. Patel, T. T. Nguyen, J. Lee, C. Choi, P. Bhatnagar, et al. (2025). 2D-SnS-Embedded Schottky Device with Neurotransmitter-Like Functionality Produced Using Proximity Vapor Transfer Method for Photonic Neurocomputing. *Adv. Mater.* 37, 2411420. Available at: https://doi.org/10.1002/adma.202411420.

Nguyen Thanh Tien, Pham Thi Bich Thao, Nguyen Thi Han, and Vo Khuong Dien (2025). Symmetry-driven valleytronics in the single-layer tin chalcogenides SnS and SnSe. *Phys. Rev. B* 109, 155416. Available at: https://doi.org/10.1103/PhysRevB.109.155416.

Nicolas Gauriot, Raj Pandya, Jack Alexander-Webber, and Akshay Rao (2025). Isolation and characterisation of monolayer phosphorene analogues. *J. Phys.: Condens. Matter* 37, 03LT01. Available at: https://iopscience.iop.org/article/10.1088/1361-648X/ad81a1#cmad81a1s6.

R. Moqbel, R. Nanae, S. Kitamura, M.-H. Lee, Y.-W. Lan, C.-C. Lee, et al. (2024). Giant Second-Order Nonlinearity and Anisotropy of Large-Sized Few-Layer SnS with Ferroelectric Stacking. *Adv. Opt. Mater.* 12, 2400355. Available at: https://doi.org/10.1002/adom.202400355.

R. Nanae, S. Kitamura, Y.-R. Chang, K. Kanahashi, T. Nishimura, R. Moqbel, et al. (2024). Bulk Photovoltaic Effect in Single Ferroelectric Domain of SnS Crystal and Control of Local

Polarization by Strain. *Adv. Funct. Mater.* 34, 2406140. Available at: https://doi.org/10.1002/adfm.202406140.

Rodin, A. S., Gomes, L. C., Carvalho, A., and Castro Neto, A. H. (2016). Valley physics in tin (II) sulfide. *Phys. Rev. B* 93, 045431. doi: 10.1103/PhysRevB.93.045431

Sarkar, A. S., Konidakis, I., Gagaoudakis, E., Maragkakis, G. M., Psilodimitrakopoulos, S., Katerinopoulou, D., et al. (2023). Liquid Phase Isolation of SnS Monolayers with Enhanced Optoelectronic Properties. *Adv. Sci.* 10, 2201842. doi: 10.1002/advs.202201842

Sarkar, A. S., Kumari, A., Anchala, Nakka, N., Ray, R., Stratakis, E., et al. (2021). Excitation dependent photoluminescence from quantum confined ultrasmall SnS sheets. *Appl. Phys. Lett.* 119, 241902. Available at: 10.1063/5.0062372.

Sarkar, A. S., Mushtaq, A., Kushavah, D., and Pal, S. K. (2020). Liquid exfoliation of electronic grade ultrathin tin(II) sulfide (SnS) with intriguing optical response. *npj 2D Mater. Appl.* 4, 1. doi: 10.1038/s41699-019-0135-1

Sarkar, A. S., and Stratakis, E. (2020). Recent Advances in 2D Metal Monochalcogenides. *Adv. Sci.* 7, 1–36. doi: 10.1002/advs.202001655

Sarkar, A. S., and Stratakis, E. (2021). Dispersion behaviour of two dimensional monochalcogenides. *J. Colloid Interface Sci.* 594, 334–341. doi: 10.1016/j.jcis.2021.02.081

Sharma, A., Yan, H., Zhang, L., Sun, X., Liu, B., and Lu, Y. (2018). Highly Enhanced Many-Body Interactions in Anisotropic 2D Semiconductors. *Acc. Chem. Res.* 51, 1164–1173. doi: 10.1021/acs.accounts.7b00504

Skelton, J. M., Burton, L. A., Oba, F., and Walsh, A. (2017). Metastable cubic tin sulfide: A novel phonon-stable chiral semiconductor. *APL Mater.* 5, 036101. doi: 10.1063/1.4977868

Sutter, E., Kisslinger, K., Unocic, R. R., Burns, K., Hachtel, J., and Sutter, P. (2024). Photonics in Multimaterial Lateral Heterostructures Combining Group IV Chalcogenide van der Waals Semiconductors. *Small* 20, 2307372. doi: 10.1002/smll.202307372

Sutter, E., Wang, J., and Sutter, P. (2020). Surface Passivation by Excess Sulfur for Controlled Synthesis of Large, Thin SnS Flakes. *Chem. Mater.* 32, 8034–8042. doi: 10.1021/acs.chemmater.0c03297

Tao Zhang, Chao Fan, Lingxiang Hu, Fei Zhuge, Xinhua Pan, and Zhizhen Ye (2024). A Reconfigurable All-Optical-Controlled Synaptic Device for Neuromorphic Computing Applications. *ACS Nano* 18, 16236–16247. Available at: https://pubs.acs.org/doi/10.1021/acsnano.4c02278.

Tian, Z., Guo, C., Zhao, M., Li, R., and Xue, J. (2017). Two-Dimensional SnS: A Phosphorene Analogue with Strong In-Plane Electronic Anisotropy. *ACS Nano* 11, 2219–2226. doi: 10.1021/acsnano.6b08704

Verma, A., Soni, A., Sarkar, A. S., and Pal, S. K. (2023). Defect-mediated saturable absorption and carrier dynamics in tin (II) monosulfide quantum dots. *Opt. Lett.* 48, 4641–4644. Available at: 10.1364/ol.498545.

Vidal, J., Lany, S., D'Avezac, M., Zunger, A., Zakutayev, A., Francis, J., et al. (2012). Band-structure, optical properties, and defect physics of the photovoltaic semiconductor SnS. *Appl. Phys. Lett.* 100, 032104. doi: 10.1063/1.3675880

Wang, C., Zhou, Z., and Gao, L. (2024). Two-Dimensional van der Waals Superconductor Heterostructures: Josephson Junctions and Beyond. *Precis. Chem* 2, 273–281. doi: 10.1021/prechem.3c00126

Wang, H., and Qian, X. (2017). Two-dimensional multiferroics in monolayer group IV monochalcogenides. *2D Mater.* 4, 015042. doi: 10.1088/2053-1583/4/1/015042

Xia, F., Wang, H., Hwang, J. C. M., Neto, A. H. C., and Yang, L. (2019). Black phosphorus and its isoelectronic materials. *Nat. Rev. Phys.* 1, 306–317. Available at: 10.1038/s42254-019-0043-5.

Y. Liu, and C. Huang (2025). Recent Advances in Metal Chalcogenide Catalysts for Electrochemical Carbon Dioxide Reduction. *Adv. Funct. Mater.*, 2505245. Available at: https://doi.org/10.1002/adfm.202505245.

Yichen Liu, Qingxiao Meng, Pezhman Mahmoudi, Ziyi Wang, Ji Zhang, Jack Yang, et al. (2024). Advancing Superconductivity with Interface Engineering. *Adv. Mater.* 36, 2405009. Available at: https://doi.org/10.1002/adma.202405009.

Yixuan Hu, Shulin Bai, Yi Wen, Dongrui Liu, Tao Hong, Shan Liu, et al. (2024). Stepwise Optimization of Thermoelectric Performance in n-Type SnS. *Adv. Funct. Mater.* 35, 2414881. Available at: https://doi.org/10.1002/adfm.202414881.

Yoo, C., Adepu, V., Han, S. S., Kim, J. H., Shin, J. C., Cao, J., et al. (2023). Low-Temperature Centimeter-Scale Growth of Layered 2D SnS for Piezoelectric Kirigami Devices. *ACS Nano* 17, 20680–20688. doi: 10.1021/acsnano.3c08826

Y.-R. Chang, R. Nanae, S. Kitamura, T. Nishimura, H. Wang, Y. Xiang, et al. (2023). Shift-Current Photovoltaics Based on aNon-Centrosymmetric Phase in In-Plane Ferroelectric SnS. *Adv. Mater.* 35, 2301172. Available at: https://doi.org/10.1002/adma.202301172.

Zhang, X., Shi, Y., Shi, Z., Xia, H., Ma, M., Wang, Y., et al. (2023). High-Pressure Synthesis of Single-Crystalline SnS Nanoribbons. *Nano Lett* 23. doi: 10.1021/acs.nanolett.3c01879

Zhao, X., Li, Z., Wu, S., Lu, M., Xie, X., Zhan, D., et al. (2024). Raman Spectroscopy Application in Anisotropic 2D Materials. *Adv. Electron. Mater.* 10, 2300610. doi: 10.1002/aelm.202300610

Zhitao Lin, Xianguang Yang, Junda He, Ning Dong, and Baojun Li (2025). Structural and optoelectronic characterization of anisotropic two-dimensional materials and applications in polarization-sensitive photodetectors. *Appl. Phys. Rev.* 12, 011301. Available at: https://doi.org/10.1063/5.0226193.

Zhu, M., Zhong, M., Guo, X., Wang, Y., Chen, Z., Huang, H., et al. (2021). Efficient and Anisotropic Second Harmonic Generation in Few-Layer SnS Film. *Adv. Opt. Mater.* 9, 2101200. doi: 10.1002/adom.202101200

Zi, Y., Zhu, J., Hu, L., Wang, M., and Huang, W. (2022). Nanoengineering of Tin Monosulfide (SnS)-Based Structures for Emerging Applications. *Small Sci.* 2, 2100098. doi: 10.1002/smsc.202100098

Zou, J., Lee, C. Y., and Wallace, G. G. (2021). Boosting Formate Production from CO2 at High Current Densities Over a Wide Electrochemical Potential Window on a SnS Catalyst. *Adv. Sci.* 8, 2004521. doi: 10.1002/advs.202004521